\documentstyle[12pt]{article}
\begin{document}


\renewcommand{\thesection}{\arabic{section}}
\renewcommand{\theequation}{\arabic{equation}}
\renewcommand {\c}  {\'{c}}
\newcommand {\cc} {\v{c}}
\newcommand {\s}  {\v{s}}
\newcommand {\CC} {\v{C}}
\newcommand {\C}  {\'{C}}
\newcommand {\Z}  {\v{Z}}

\baselineskip=24pt


\begin{center}
{\bf   A multispecies   Calogero model}
 
\bigskip

S.Meljanac$^{a}$ {\footnote{e-mail: meljanac@thphys.irb.hr}}, 
 M.Milekovi\'{c} $^{b}$ {\footnote{e-mail: marijan@phy.hr }}
  and  A. Samsarov$^{a}${ \footnote{e-mail:andjelo.samsarov@irb.hr}}\\

$^{a}$ Rudjer Bo\v{s}kovi\'c Institute, Bijeni\v cka  c.54, HR-10002 Zagreb,
Croatia\\[3mm] 

$^{b}$ Theoretical Physics Department, Faculty of Science, P.O.B. 331,
 Bijeni\v{c}ka c.32,
\\ HR-10002 Zagreb, Croatia \\[3mm]

\bigskip

\end{center}
\setcounter{page}{1}
\bigskip

\begin{center}
\bf Abstract
\end{center}
We study a multispecies one-dimensional Calogero model with two- and three-body interactions.
Using an algebraic approach (Fock space analysis), we construct ladder operators and find infinitely many, 
but not all, exact eigenstates of the model  Hamiltonian. Besides the ground state energy, we deduce 
energies of the excited states. It turns out that the  spectrum is linear in quantum numbers and
 that the higher-energy levels are degenerate. The dynamical symmetry responsible for degeneracy is $SU(2)$. 
We also find the universal critical point at which the model exhibits singular behaviour.
Finally, we make  contact with some special cases mentioned in the literature.

\bigskip
PACS number(s): 03.65.Fd, 03.65.Sq, 05.30.Pr


\newpage




\section {Introduction}
The ordinary Calogero [1] model describes $N$ indistinguishable particles on the line which
interact through an inverse-square two-body interaction and are subjected to a common 
confining harmonic force. The model is completely integrable in both the classical and quantum case 
[2], the spectrum is known and the wave functions are given implicitly. The model and its
various descendants (also known as Calogero-Sutherland-Moser systems [3]) are connected with 
a number of physical problems, ranging from condensed matter
physics [4] to gravity and black hole physics [5]. The algebraic structure of the Calogero
model, studied earlier using group theoretical methods [2,6], has recently
been reconsidered by a number of authors in the framework of the $S_N$ (permutational) algebra 
[7]. This operator approach is considerably simpler than the original one, yields an explicit
expression for the wavefunctions  and emphasizes the interpretation in terms of generalized
statistics [8], especially Haldane's exclusion statistics [9]. 
In Haldane's formulation there is the 
possibility of having particles of different species with a mutual statistical coupling parameter depending
on the $i^{th}$ and $j^{th}$ species coupled. On the level of the Calogero model, this corresponds to
the generalization of the ordinary 1D Calogero model with identical particles to the 1D Calogero model with 
non-identical particles. This can be done by allowing particles to have different  masses and different 
 couplings to each other. In this way we obtain a 1D multispecies Calogero model. Very little is known
about its spectra and wavefunctions  [10-13]. 

In the present  Letter, which is in a sense a continuation of our
investigation of the ordinary Calogero model [14], we use an algebraic (operator) method to find some of the salient features
of the  multispecies Calogero model on the line with two- and three-body interactions.  
In Section 2 we present the $S_N$ invariant  Hamiltonian $H$ of the model, together with its ground state 
wavefunction and ground state energy. After performing a certain similarity transformation, we get
a  much simpler 
Hamiltonian $\tilde{H}$, which we separate into  parts describing the center-of-mass motion and the
relative motion 
of particles. We express $\tilde{H}$ in terms of generators of the $SU(1,1)$ algebra. All analysis is made 
in Hilbert space.
Section 3 contains our most important results. By applying Fock space analysis, we find some of the  
excited states of $\tilde{H}$, their energies and degeneracies.  Closer inspection of the Fock space that 
corresponds to the relative motion of particles reveals  the existence of the universal critical point at which 
the system exhibits singular behaviour. This result generalizes that mentioned in [14]. We also establish the 
 necessary conditions for the equivalence of the two multispecies Calogero models.
In Section 4 we briefly repeat the main points
of the paper and make  contact with the models studied in [10-13]. We particularly discuss the necessary
conditions for vanishing of the three-body interaction in the starting Hamiltonian $H$.



\section {A multispecies Calogero model with a three-body interaction}

Let us consider the most general Calogero type ground state  for the 
$N$-body quantum mechanical problem on the line ($\hbar  =1$):
\begin{equation}
 \Psi_0 (x_1,x_2,\cdots x_N)=\Delta e^ {-\frac{\omega}{2}
\sum^{N}_{i=1} m_i x_i^2},
\end{equation}
where the prefactor $\Delta$ is given by
$$
\Delta =\prod_{i<j} (x_i -x_j)^{\nu_{ij}}, \qquad \nu_{ij}=\nu_{ji} ,\qquad i,j=1,2 \cdots
N
$$
and $\nu_{ij}$ are symmetric statistical parameters between particles $(i,j)$. The harmonic frequency 
$\omega$ is taken to be the same for all particles. Masses of the particles ( $m_i$ ) are, in general, not equal.

The Hamiltonian which possesses the above state (1) as the ground state is \\
( $p_i=- i \frac{\partial}{\partial x_i}$ )

$$
H=\sum_{i=1}^N \frac{p_i^2}{2m_i} + \frac{\omega^2}{2}\sum_{i=1}^N m_ix_i^2 +
\frac{1}{4}\sum_{i\neq j }\frac{\nu_{ij}(\nu_{ij}-1)}{(x_i-x_j)^2} (\frac{1}{m_i} +
\frac{1}{m_j})  
$$

\begin{equation}
+\frac{1}{2}\sum_{i,j,k \neq}\frac{\nu_{ij}\nu_{jk}}{m_j(x_j-x_i)
(x_j-x_k)} ,
\end{equation}

$$
H \Psi_0 (x_1,x_2,\cdots x_N)=E_0 \Psi_0 (x_1,x_2,\cdots x_N),
$$

$$
E_0=\omega (\frac{N}{2}+ \frac{1}{2}\sum_{i\neq j}\nu_{ij}).
$$
The symbol $(i,j,k \neq ) $ in the  last term denotes the summation over 
all triples
of mutually distinct indices.\\
The Hamiltonian in Eq.(2) 
is invariant under the  group of permutation of $N$ elements, $S_N$.
The elementary generators $K_{ij}$ of the symmetry group $S_N$ exchange
labels $i$ and $j$ in $ all$  $quantities$, according to the rules:
$$
K_{ij}x_j=x_iK_{ij},\; K_{ij}m_j=m_iK_{ij}, \; K_{ij}\nu_{jl}= \nu_{il}K_{ij},
$$
$$
K_{ij}=K_{ji},\; (K_{ij})^2=1, 
$$
$$
 K_{ij}K_{jl}=K_{jl}K_{il}=K_{il}K_{ij},\;{\rm for}\;i\neq j,\;
i\neq l,\;j\neq l.
$$
 It follows that $ K_{ij}HK_{ij}=H$, i.e. $[H,K_{ij}]=0$, $\forall i,j$.\\
In a sense, one may think of (2) as a smooth, sufficiently small deformation
 of the ordinary Calogero model.

A few additional remarks concerning the Hamiltonian (2) are in order.\\
(1) It describes distinguishable particles on the line, interacting with 
harmonic, two-body and three-body potentials. As far as we know, it is the first time that such Hamiltonian is
considered in full generality. An earlier attempt to solve the similar, but less general Hamiltonian 
( with all masses $m_i$ equal) can be find in [11] (see also Ref.[13]).\\ 
(2) The asymptotic behaviour of its eigenstates should be 
$\Psi \propto (x_i -x_j)^{\nu_{ij}}$ as \\$(x_i -x_j) \rightarrow 0$.\\
(3) A well-known stability condition  demands that the two-body 
couplings \\ $\nu_{ij}(\nu_{ij} - 1)$ should be greater 
than $-\frac{1}{4}$, $\forall i,j$.\\
(4) Setting $\nu_{ij}=\nu$, $\forall i,j$ and $m_i = m$, $\forall i$, we recover 
the ordinary N-body Calogero model [1]. In that case,  owing  to the identity $ \sum_{cycl.} \frac{1}{(x_i-x_j)(x_i-x_k)}=0$
  which holds in 1D,  the three-body term in Eq.(2) trivially vanishes.
When $\nu_{ij}$ are mutually different  but $m_i = m$, $\forall i$,
we recover the model treated in [11,13]. Finally, 
when $\nu_{ij}=\alpha m_i m_j$, $\alpha$ being constant, we obtain the model 
 mentioned in [10,12]. We comment on these cases in  Section 4.\\
In the following we analyse the most general case, namely the Hamiltonian of Eq.(2)
without restrictions $(4)$, given above.

Let us perform the non-unitary transformation on 
 $ \Psi_0 (x_1,x_2,\cdots x_N)$:
\begin{equation}
\tilde{\Psi}_{0}(x_1,x_2,\cdots x_N)=
\Delta^{-1}\Psi_0 (x_1,x_2,\cdots x_N)=e^{-\frac{\omega}{2}
\sum^{N}_{i=1} m_i x_i^2} .
\end{equation}
It generates a similarity transformation which leads to  another $S_N$ invariant 
(but non-Hermitean)  Hamiltonian $\tilde{H}$:
$$
\tilde{H}=\Delta^{-1} H \Delta ,
$$
$$
\tilde{H} \tilde{\Psi}_{0}(x_1,x_2,\cdots x_N) = E_0 
\tilde{\Psi}_{0}(x_1,x_2,\cdots x_N),
$$
with $E_0$ given in Eq.(2).\\
We find $\tilde{H}$ as
\begin{equation}
\tilde{H} = -\frac{1}{2}\sum_{i=1}^N \frac{1}{m_i}\frac{\partial^2}{\partial x_i^2} 
+ \frac{\omega^2}{2}\sum_{i=1}^N m_i x_i^2 - \frac{1}{2}\sum_{i\neq j}\frac{\nu_{ij}}{(x_i - x_j)}
(\frac{1}{m_i}\frac{\partial}{\partial x_i} - \frac{1}{m_j}\frac{\partial}{\partial  x_j}) .
\end{equation}
Notice that in Eq.(4)  two-body and  three-body interactions apparently disappeared
but they are hidden in the last term of Eq. (4).\\
It is convenient to introduce the  variables $(X,\xi_i)$ 
\begin{equation}
X = \frac{\sum_{i=1}^N m_i x_i}{\sum_{i=1}^N m_i} \equiv 
\frac{1}{M}{\sum_{i=1}^N m_i
x_i} \quad ,\qquad \xi_i=x_i -X, \quad i=1,2,\cdots N,
\end{equation}
and the  linear combinations of derivatives $(\frac{\partial}{\partial \xi_i},
\frac{\partial}{\partial X})$
\begin{equation}
\frac{\partial}{\partial \xi_i}=\frac{\partial }{\partial x_i}-\frac{m_i}{M}\frac{\partial }
{\partial X}\quad, \qquad 
\frac{\partial }{\partial X}=\sum_{i=1}^N \frac{\partial}{\partial x_i},\quad i=1,2,\cdots N.
\end{equation}
Note that the variables $\xi_i$ ( as well as  $\frac{\partial }{\partial \xi_i}$ ) are not 
linearly independent, i.e.\\ $\sum_{i=1}^Nm_i\xi_i = \sum_{i=1}^N\frac{\partial }{\partial \xi_i}=0$.\\
In terms of the variables just introduced, the Hamiltonian $\tilde{H}$, Eq.(4), separates
into parts which describe its center-of-mass motion (CM) and its relative motion (R),
namely $\tilde{H}=\tilde{H}_{CM} + \tilde{H}_{R}$, with
$$
\tilde{H}_{CM}= -\frac{1}{2M}\frac{\partial ^2}{\partial X^2} + \frac{1}{2}M \omega^2 X^2,
$$
\begin{equation}
\tilde{H}_{R}= -\frac{1}{2}\sum_{i=1}^N\frac{1}{m_i}\frac{\partial ^2}{ \partial \xi_i^2}
+   \frac{1}{2}\omega ^2 \sum_{i=1}^N m_i\xi_i^2
-\frac{1}{2}\sum_{i\neq j}\frac{\nu_{ij}}{(\xi_i - \xi_j)}
( \frac{1}{m_i}\frac{\partial }{\partial \xi_i}  - \frac{1}{m_j}\frac{\partial }{\partial \xi_j} ).
\end{equation}
The Hamiltonian $H$, Eq.(2), can also be decomposed into $H_{CM}$ and $H_{R}$. \\
The wave function (3) separates as 
$$
\tilde{\Psi}_{0}(x_1,x_2,\cdots x_N)
=\tilde{\Psi}_{0}(X)\tilde{\Psi}_{0}(\xi_1, \xi_2 \cdots \xi_N)=
e^{-\frac{M\omega}{2}X^2} e^{-\frac{\omega}{2}\sum^{N}_{i=1} m_i \xi_i^2} .
$$
The ground state energy $E_0$ splits into the energy of CM 
( $E_{0CM}=\frac{1}{2}\omega$ ) and the energy
of relative motion ($E_{0R}= \frac{N-1}{2}\omega+ \frac{1}{2}\omega
\sum_{i\neq j}\nu_{ij}$). \\
We define the set of operators $\{T_{+}, T_{-}, T_{0}\}$ as
$$
T_{+}=\frac{1}{2}\sum_{i=1}^N m_i x_i^2 ,
$$
$$
T_{-}= \frac{1}{2}\sum_{i=1}^N \frac{1}{m_i}\frac{\partial ^2}{\partial x_i^2} 
+ \frac{1}{2}\sum_{i\neq j}\frac{\nu_{ij}}{(x_i - x_j)}
(\frac{1}{m_i}\frac{\partial }{\partial x_i} - \frac{1}{m_j}\frac{\partial }{\partial x_j})
,
$$

\begin{equation}
T_0=\frac{1}{2}( \sum_{i=1}^N x_i\frac{\partial }{\partial x_i} + \frac{E_0}{\omega} )
.
\end{equation}
Using Eqs.(5) and (6) one can easily show that these operators also split as\\ $T_{\pm,0}=
T_{\pm,0_{(CM)}} + T_{\pm,0_{(R)}}$.

Operators  $\{T_{+}, T_{-}, T_{0}\}$ satisfy the $SU(1,1)$ algebra:
$$ 
[T_{-},T_{+}]=2 T_0 \quad , \qquad [T_{0},T_{\pm}]=\pm T_{\pm}.
$$
In terms of these operators, the Hamiltonian (4) reads $\tilde{H}=\omega^2 T_+
- T_-$.


\section {Ladder operators and Fock space representation}

Now we introduce pairs of creation and annihilation operators $\{ A_1^{+},A_1^{-}\}$ 
and $\{A_2^{+}, A_2^{-}\}$:
$$
A_1^{\pm}=\frac{1}{\sqrt 2} ( \sqrt{ M \omega}X \mp \frac{1}{\sqrt{ M
\omega}}\frac{\partial }{\partial X}) ,
$$
\begin{equation}
A_2^{\pm}=\frac{1}{2} ( \omega T_+ + \frac{1}{\omega} T_- )\mp T_0 ,
\end{equation}
which satisfy the following commutation relations:
$$
[A_1^{-}, A_1^{+}] = 1, \qquad [A_2^{-}, A_2^{+}] = \frac{1}{\omega} 
\tilde {H},
$$
$$
[A_1^{-},A_2^{-}]=[A_1^{+},A_2^{+}]=0,\qquad  [A_1^{-},A_2^{+}]=A_1^{+}, \qquad [A_1^{+},A_2^{-}]= - A_1^{-},
$$
\begin{equation}
[\tilde {H},A_1^{\pm}] = \pm \omega A_1^{\pm}, 
\end{equation}
$$
[\tilde {H},A_2^{\pm}] = \pm 2 \omega A_2^{\pm} .
$$
They act on the Fock vacuum $|\tilde {0}\rangle \propto 
\tilde{\Psi}_{0}(x_1,x_2,\cdots x_N)$ as
$$
A_1^{-}|\tilde {0}\rangle = A_2^{-}|\tilde {0}\rangle =0, \qquad 
\langle\tilde {0}|\tilde {0}\rangle =1.
$$
The excited states are built as
\begin{equation}
|n_1, n_2\rangle \propto A_1^{+ n_1} A_2^{+ n_2}|\tilde {0}\rangle ,
\qquad \forall n_1,n_2 =0,1,\cdots
\end{equation}
 The repeated action of the operators $A_1^{+}$ ( $A_2^{+}$ ) on the vacuum 
$|\tilde {0}\rangle $
reproduces in the coordinate representation Hermite polynomials ( hypergeometric function ),
 respectively.

The states $|n_1, n_2\rangle $ are eigenstates of $\tilde{H} $, Eq.(4), with 
the eigenvalues\\ ( cf. Eq.(10) )
\begin{equation}
E_{n_1,n_2}=\omega ( n_1 + 2 n_2 ) + E_0 .
\end{equation}
The energy spectrum is $ linear $ in quantum numbers $n_1,n_2$. This result is
universal, i.e. it holds for all parameters $m_i$ and $\nu_{ij}$ in the
Hamiltonian $\tilde {H}$ (or $H$ ). Notice that the energy of the ground state  and 
the energy of the first
excited state are non-degenerate whereas  the higher energy levels are degenerate.
The structure of degeneracy is as follows:
$$
\begin{tabular}{|c|c|c|c|} \hline
$n_1$ & $n_2$ & $n=n_1+2 n_2$ &  Degenerate states     \\ \hline \hline
0   & 0   &      0      &    $  |\tilde{0}\rangle $         \\ \hline
1   & 0   &      1      &    $A_1^+  |\tilde{0}\rangle $     \\ \hline \hline
2   & 0   &      2      &    $A_1^{+ 2}  |\tilde{0}\rangle $  \\ 
0   & 1   &      2      &    $A_2^+  |\tilde{0}\rangle $       \\ \hline \hline
1   & 1   &      3      &    $ A_1^+  A_2^+  |\tilde{0}\rangle $ \\ 
3   & 0   &      3      &    $A_1^{+ 3}  |\tilde{0}\rangle $      \\ \hline \hline
0   & 2   &      4      &    $A_2^{+ 2}  |\tilde{0}\rangle $      \\ 
2   & 1   &      4      &    $A_1^{+ 2} A_2^+ |\tilde{0}\rangle $   \\ 
4   & 0   &      4      &    $A_1^{+ 4}  |\tilde{0}\rangle $      \\ \hline \hline
5   & 0   &      5      &    $A_1^{+ 5}  |\tilde{0}\rangle $      \\ 
3   & 1   &      5      &    $A_1^{+ 3} A_2^+ |\tilde{0}\rangle $     \\ 
1   & 2   &      5      &    $A_1^{+ } A_2^{+ 2} |\tilde{0}\rangle $   \\ \hline \hline
$\vdots$ & $\vdots$ & $\vdots$ & $\vdots$ \\ \hline 

\end{tabular}
$$
It is evident that for $n=even$, the degeneracy is $(\frac{n}{2} +1)$ and for 
$n=odd $, the degeneracy is $(\frac{n+1}{2})$.\\
In order that the two models described by Hamiltonian (2), 
with statistical parameters $\nu_{ij}$ and  $\nu'_{ij}$, have the same tower of states (11) and
the same spectrum (12), the  necessary conditions are $\sum_{i<j}\nu_{ij}=  
\sum_{i<j}\nu'_{ij}$, $\omega =\omega'$ and the number of particles should be the same, i.e.
$N=N'$.\\
The dynamical symmetry algebra of the  model is defined as maximal algebra commuting with the Hamiltonian 
$ \tilde {H}$. 
The  dynamical symmetry of the ordinary Calogero model is 
complicated polynomial algebra denoted by $C_N(\nu)$ in [15].
In our  case,   owing to the fact that $ \tilde {H}$ (10) can be rewritten in terms of two independent,
uncoupled oscillators (see bellow Eqs.(13) and (14)), this polynomial algebra can be linearized to the ordinary
$SU(2)$ algebra. This is the minimal symmetry that remains in the generic case, i.e. for general $\nu_{ij}$ and
$m_i$. In fact, this is the same  dynamical symmetry underlying the two-body Calogero model [15,16].

We point out that one can construct the creation and annihilation operators 
$\{B_{2}^{+},B_{2}^{-}\}$:  
\begin{equation}
B_{2}^{\pm} = {A_{2}}^{\pm} - \frac{1}{2}{A_{1}^{\pm}}^{2} .
\end{equation}
In terms of $A_{1}^{\pm}$ and $B_{2}^{\pm}$, the above-mentioned $SU(2)$
algebra ($[J_{+},J_{-}]=2J_{0}$, $[J_{0},J_{\pm}]=\pm J_{\pm}$) is generated by 
$$
J_{+} = A_{1}^{+ 2}B_{2}^{-}\frac{1}{\sqrt {2(\hat{N}_2 - 1 + \frac{E_{0R}}{\omega})(\hat{N}_1 +1)}},
$$
$$
J_{-} = B_{2}^{+} A_{1}^{-2} \frac{1}{\sqrt {2(\hat{N}_2  + \frac{E_{0R}}{\omega})(\hat{N}_1 -1)}},
$$
$$
J_{0} = \frac{1}{4} (\hat{N}_1 - 2\hat{N}_2).
$$
Here, $\hat{N}_1$ and $\hat{N}_2$ are number operators counting $A_1$- and $B_2$- modes, respectively.

 The benefit of  the construction (13) is that   
the operators $A_{1}^{\pm}$, corresponding to the CM motion,
  decouple completely (cf. Eq.(10)), i.e.
$$
[A_{1}^{\pm},B_{2}^{\mp}] = 0.
$$
Hence, we get
$$
\tilde{H}_{CM}=\frac{1}{2}\omega \{A_{1}^{-},A_{1}^{+}\}_+ \equiv \omega (\hat{N}_1 + \frac{E_{0CM}}{\omega}),
$$
$$
\tilde{H}_{R}= \omega [B_{2}^{-},B_{2}^{+}] \equiv \omega (2 \hat{N}_2 + \frac{E_{0R}}{\omega}),
$$
\begin{equation}
[\tilde{H}_{R},B_{2}^{\pm}] = \pm 2 \omega B_{2}^{\pm} .
\end{equation}
The Fock space now splits  into the CM-Fock space, spanned by 
$A_{1}^{+ n_1}|\tilde{0}\rangle_{CM} $, and the R-Fock space, spanned by
$B_{2}^{+ n_2}|\tilde{0}\rangle_{R} $, where 
$|\tilde{0}\rangle_{CM}\propto \tilde{\Psi}_{0}(X)$ and 
$ |\tilde{0}\rangle_{R} \propto \tilde{\Psi}_{0} (\xi_1, \xi_2 \cdots \xi_N)$.\\

At this point it is useful to make a contact with the Fock space of the ordinary Calogero model. 
In the ordinary Calogero model the $S_N$-symmetric subspace of the whole Fock space  is spanned by the   states 
$ {\cal {A}}_1^{+ n_1}|\tilde{0}\rangle_{CM}$ and  
$({\cal {B}}_2^{+ n_2}\cdots  {\cal {B}}_N^{+ n_N}) |\tilde{0}\rangle_{R}$ [7,16], where 
$$
{\cal {A}}_1=\sum_{i=1}^N a_i ,  \qquad {\cal {B}}_k=\sum_{i=1}^N (a_i - \frac{1}{N}{\cal{A}}_1)^k , 
\qquad  2 \leq k \leq N .
$$
The so-called Dunkl-Polychronakos operators $a_i$  satisfy the algebra [7]
\begin{equation}
[a_i,a_j^{\dagger}]= \delta_{ij}(1+ \nu \sum_{k=1}^N K_{ik}) - \nu K_{ij} ,
\end{equation}
$$
[a_i^{\dagger},a_j^{\dagger}] =[a_i,a_j]=0.
$$
Our operators $A_{1}$ and $B_{2}$ correspond exactly to the operators ${\cal {A}}_1$ and ${\cal {B}}_2$ in the 
ordinary Calogero model. Within our algebraic treatment, we are unable to construct the eigenstates of (2)
 which correspond to the  Calogero-states $({\cal {B}}_3^{+ n_3}\cdots  {\cal {B}}_N^{+ n_N}) |\tilde{0}\rangle_{R}$ .

Reducing the problem (2) to the (4) and (14), i.e. 
$H\rightarrow \tilde{H}\rightarrow \tilde{H}_{R}$, has an interesting consequence,
namely the existence of the universal critical point, defined by the null-vector
\begin{equation}
\frac{_R\langle \tilde{0} | B_{2}^{-} B_{2}^{+}| \tilde{0} \rangle_R}{_R\langle \tilde{0} |\tilde{0} \rangle_R} = 
\frac{N-1}{2} + \frac{1}{2}\sum_{i\neq j } \nu_{ij}  = 0 .
\end{equation}
The above relation (16) follows directly from Eq.(14) by demanding \\
$_R\langle \tilde{0} |\tilde{H}_{R} | \tilde{0} \rangle_R= E_{0R}=0$. More generally, from Eq.(14) (see also
Ref.[17] ) immediately follows that
$$
B_{2}^{+} B_{2}^{-}= \hat{N}_2 (\hat{N}_2 - 1 + \frac{E_{0R}}{\omega})\equiv \phi (\hat{N}_2),
$$
$$
B_{2}^{-} B_{2}^{+}= (\hat{N}_2 + 1)(\hat{N}_2  + \frac{E_{0R}}{\omega})\equiv \phi (\hat{N}_2 +1),
$$
\begin{equation}
\frac{_R\langle \tilde{0} | B_{2}^{- m} B_{2}^{+ m}| \tilde{0} \rangle_R}{_R\langle \tilde{0} |\tilde{0}
\rangle_R}= \prod_{k=1}^{m} \phi (k).
\end{equation}
Since Eq.(16) implies that $\phi (1)= \frac{E_{0R}}{\omega}= 0$, it is obvious that the critical point (Eq.(16)) is
unique, i.e. there are no similar critical points when norms of states involving higher powers of the operators 
$B_2$ are involved.

At the critical point  the system described by  $\tilde{H}_{R}$  collapses completely. 
This means that the relative coordinates, 
the relative momenta and the relative energy  are all zero at this critical 
point. There survives only one oscillator, describing the motion of the
centre-of-mass.  This singular behaviour was first noticed in [14] for the case 
$ \nu_{ij}=\nu$ and $m_i=m$.
Of course, for the initial Hamiltonian $H$, which is not unitary 
( i.e. physically ) equivalent to  $\tilde{H}$, this corresponds to
 some $\nu_{ij} < 0$ and the norm of the wave function (1) blows up at the critical point. 
For $\nu_{ij}$ negative but greater than the critical values (16), the wave function is singular 
at coincidence points but still quadratically integrable.  



\section {Discussion and outlook}
In this Letter we have studied the most general multispecies Calogero model on the line, Eq.(2)
, with a three-body
interaction and an extended $S_N$ invariance. By applying the similarity transformation $\Delta $, we have 
obtained the   
Hamiltonian (4), on which  we have performed the   Fock space analysis (9,10) and found some of its  
(but not all) excited states, Eq.(11), and their energies $E_{n_1,n_2}$, (12). 
It turns out that the energy (12) is $linear$ in quantum numbers $n_1$ and $n_2$ and there is 
a dynamical $SU(2)$ symmetry responsible for the degeneracy of higher-energy levels  with 
$n=n_1+ 2 n_2 \geq 2$. By splitting the Fock space into the CM-Fock space and the R-Fock space
 (14), we have 
detected  the universal critical point (16) at which the system exhibits singular behaviour.

To conclude this analysis, let us consider  the last term in (2) more closely, 
namely

\begin{equation}
\frac{1}{2}\sum_{i,j,k \neq}(\frac{\nu_{ij}\nu_{jk}}{m_j})\frac{1}{(x_j-x_i)
(x_j-x_k)}.
\end{equation}

If we put  $m_j=m=const.$ in (18), $\forall j$, symmetrize under the cyclic exchange of the
indices ($i\rightarrow j \rightarrow k \rightarrow i $) and reduce the sum to a common 
denominator using the identity
$$
\sum_{cycl.} \frac{1}{(x_i-x_j)(x_i-x_k)}=0,
$$
we obtain that the necessary condition for vanishing of the three-body interaction is
$\nu_{ij}=\nu = const.$, $\forall i,j$. In this way, the problem (2) is reduced to the
ordinary N-body Calogero model with two-body interactions only [1].\\
For the general $\nu_{ij}$ and $m_j$, the above procedure yields the following necessary 
conditions for the absence of the three-body interaction (18): 
\begin{equation}
\frac{\nu_{ij}\nu_{jk}}{m_j}=\frac{\nu_{jk}\nu_{ki}}{m_k}=\frac{\nu_{ki}\nu_{ij}}{m_i} .
\qquad \forall (i,j,k)
\end{equation}
The unique solution of these conditions is $\nu_{ij}=\alpha \, m_i\, m_j$,  $\alpha $ being 
some universal constant. This particular connection between masses and interaction
parameters was also displayed in [10,12]. In [12], the condition (19) arose from the demand that the
asymptotic Bethe ansatz should be  applicable to the ground state of a multispecies many-body quantum
system obeying mutual statistics, while in [10] its origin 
was not obvious. In our approach, it has the simplest possible interpretation.\\
The results presented here are easily applied to the model with $F$ distinct families of
particles. For example, for $\alpha =1$, the ground state energy becomes
\begin{equation}
E_0=\omega (\frac{N}{2} + \sum_{a=1}^F m_a^2 \frac{N_a (N_a - 1)}{2} + \frac{1}{2}\sum_{a\neq b }^F 
m_a m_b  N_a N_b) ,
$$
$$
N= \sum_{a=1}^F N_a .
\end{equation}
Two  systems  characterized by 
$\{\omega, m_a, N_a \}$ and $\{\omega ', m_a', N_a' \}$ are identical if  $\sum N_a=\sum N_a'$,
$\omega = \omega '$ and $E_0=E_0'$. 

The open problem that still remains is the construction of (generalized) Dunkl-Polychronakos operators 
$a_i$ and $a_i^{\dagger}$ (15), which may help in finding the complete set of eigenstates of
the Hamiltonian (2) (or (4)), in one-to-one correspondence with the ordinary Calogero model.

{\bf Acknowledgment}\\
This work was supported by the Ministry of Science and Technology of the Republic of Croatia under 
contracts No. 0098003 and No. 0119261.

\newpage



\end{document}